\documentclass[a4paper,11pt]{article}
\pdfoutput=1 

\usepackage{jheppub} 

\usepackage[T1]{fontenc} 

\usepackage[multiple]{footmisc}

\usepackage{graphicx}
\usepackage{amssymb, amsmath,mathtools,amsthm}
\usepackage{hyperref}
\hypersetup{
    colorlinks=true,
    linkcolor=black,          
    citecolor=black,        
    filecolor=black,      
    urlcolor=black          
}

\newcommand{\ud}{\mathrm{d}}

\newcommand{\beq}{\begin{equation}}
\newcommand{\eeq}{\end{equation}}

\newcommand{\diff}{{\rm Diff}}

\newcommand{\vect}{{\rm Vect}}

\newcommand{\lieg}{\mathfrak{g}}

\newcommand{\ad}{{\rm ad}}
\newcommand{\Ad}{{\rm Ad}}

\newcommand{\sch}{{\rm Sch}}
\newcommand{\vir}{\mathfrak{vir}}

\newcommand{\xp}{x^+}
\newcommand{\xm}{x^-}
\newcommand{\tp}{t^+}
\newcommand{\tm}{t^-}

\newcommand{\ads}{AdS$_3$}
\newcommand{\ds}{dS$_3$}

\title{\boldmath Asymptotic symmetries of three dimensional gravity and the membrane paradigm}

\author[a]{Mariana Carrillo-Gonz\'alez}
\affiliation[a]{Center for Particle Cosmology, Department of Physics and Astronomy,\\
University of Pennsylvania, 209 S. 33rd St., Philadelphia,
PA 19104, USA}
\emailAdd{cmariana@sas.upenn.edu}

\author[b]{and Robert F. Penna}
\affiliation[b]{Center for Theoretical Physics, Department of Physics,\\
Columbia University, New York, NY 10027, USA}
\emailAdd{rp2835@columbia.edu}

\abstract{The asymptotic symmetry group of three-dimensional (anti) de Sitter space is the two dimensional conformal group with central charge $c=3\ell/2G$.  Usually the asymptotic charge algebra is derived using the symplectic structure of the bulk Einstein equations.  Here, we derive the asymptotic charge algebra by a different route.  First, we formulate the dynamics of the boundary as a 1+1-dimensional dynamical system.  Then we realize the boundary equations of motion as a Hamiltonian system on the dual Lie algebra, $\mathfrak{g}^*$, of the two-dimensional conformal group.  Finally, we use the Lie-Poisson bracket on $\mathfrak{g}^*$ to compute the asymptotic charge algebra.  This streamlines the derivation of the asymptotic charge algebra because the Lie-Poisson bracket on the boundary is significantly simpler than the symplectic structure derived from the bulk Einstein equations.  It also clarifies the analogy between the infinite dimensional symmetries of gravity and fluid dynamics.
}

\begin{document} 
\maketitle
\flushbottom

\section{Introduction}
\label{sec:intro}

The space of solutions of the Einstein equations has a huge gauge symmetry: metrics differing by small diffeomorphisms\footnote{Here ``small diffeomorphisms'' means diffeomorphisms which act trivially at asymptotic infinity.} are physically equivalent.  This introduces a great deal of complexity into the construction of a symplectic structure on this space \cite{Wald:1999wa,Barnich:2001jy}.  However, when the phase space of a dynamical system carries redundant information, it is often possible to ``quotient out'' the redundancy through the process of symplectic reduction and obtain a simpler description of the same dynamics.  An extremely simple example appears in rigid body dynamics.  In this case, the configuration space is the group $G=SO(3)$ and the canonical phase space is the cotangent bundle, $T^*G$.  However, the Euler equations of rigid body dynamics are invariant under the left action of $G$ on $T^*G\cong G\times \lieg^*$, so we can quotient out by $G$ and obtain an equivalent Hamiltonian system on $\lieg^*$, the dual of the Lie algebra\footnote{Of course, in this example the redundancy is not a gauge redundancy.}.  In this process, the canonical Poisson bracket on $T^*G$ descends to the Lie-Poisson bracket on $\lieg^*$ and the dimension of phase space is reduced by half \cite{marsden2002introduction}.

In this paper, we consider similar examples in general relativity.  Our examples are three dimensional gravity with asymptotically de Sitter (dS) and anti de Sitter (AdS) boundary conditions.  In either case, the space of physically distinct solutions of the Einstein equations is parametrized by boundary data supported at asymptotic infinity.  The projection of the Einstein equations onto asymptotic infinity gives a set of constraints on the boundary data.  We regard the constraints as equations of motion for a dynamical system supported on the 1+1-dimensional boundary.   This viewpoint is closely related to the black hole membrane paradigm, which treats the projection of Einstein's equation onto a black hole event horizon as a lower dimensional dynamical system \cite{1986bhmp.book.....T,1998PhRvD..58f4011P,2015PhRvD..91h4044P}.  Physically, the dynamics is equivalent to conservation of the boundary's Brown-York stress tensor.

We show that the dynamics so obtained can be regarded as  Hamiltonian flow on $\lieg^*$, where now $G$ is the asymptotic symmetry group of the family of spacetimes under consideration and $\lieg^*$ is the dual of its Lie algebra.   In both of our examples, $G$ is two copies of the Virasoro group with central charge fixed by the boundary equations of motion and the normalization of the Brown-York stress energy tensor.   As in the rigid body example, the Poisson bracket is the Lie-Poisson bracket and the simplicity of the Lie-Poisson bracket streamlines the computations.  The resulting formalism resembles the Hamiltonian description of the compressible Euler equations \cite{marsden1984semidirect}, and we use this analogy to clarify the relationship between conservation laws in fluid dynamics and gravity (see also \cite{Penna:2015gza,Eling:2016xlx,Eling:2016qvx,Penna:2017bdn}).

The examples in this paper are closely related to the description of three dimensional asymptotically flat gravity in \cite{Penna:2017vms}.  In the future, it would be interesting to derive the Lie-Poisson brackets appearing in all these examples from the Poisson bracket on covariant phase space \cite{Wald:1999wa,Barnich:2001jy} via symplectic reduction.  In the meantime, the subject of asymptotic symmetries remains of great current interest for its wide array of applications, ranging from consistency relations in cosmology \cite{Maldacena:2002vr,Hinterbichler:2013dpa,Berezhiani:2013ewa} to soft theorems in quantum field theory and memory effects in gravitational wave experiments (for a recent review see \cite{Strominger:2017zoo}).  The perspective we develop here can, in some cases, allow one to identify asymptotic symmetry groups and compute their charge algebras with relative ease, tools which should prove useful in this rapidly expanding subject.

\section{Virasoro algebra}
\label{sec:vir}

In this section, we collect some standard facts about the Virasoro algebra (see, e.g., \cite{khesin2008geometry,marsden2002introduction}) that will be needed in the following sections.  This establishes notation and keeps the present paper self-contained.  

The Virasoro group is a central extension of $\diff(S^1)$, the diffeomorphism group of the circle. So we start by describing $\diff(S^1)$. The elements of $\diff(S^1)$ are smooth invertible maps $S^1\rightarrow S^1$, the group multiplication is composition, and the Lie algebra is $\vect(S^1)$, the algebra of vector fields on the circle.  The Lie bracket is the usual vector field commutator.  As a vector space, $\vect(S^1)$ has a dual space, the space of linear functionals on $\vect(S^1)$.  It is convenient to identify the dual space of $\vect(S^1)$ with the space of quadratic differentials on $S^1$.  Given a quadratic differential, $u(\theta)d\theta^2$, and a vector field, $f(\theta)\partial_\theta$, we form the pairing,
\beq
\langle f(\theta)\partial_\theta,u(\theta)d\theta^2 \rangle = \int_{S^1} f(\theta) u(\theta) d\theta.
\eeq 
For each quadratic differential, this pairing defines a linear map $\vect(S^1)\rightarrow \mathbb{R}$.  So the dual of $\vect(S^1)$ becomes identified with the space of quadratic differentials.  As vector spaces, $\vect(S^1)$ and its dual are isomorphic.  However, vector fields and quadratic differentials transform differently under diffeomorphisms.  This will be important in what follows.

The Virasoro algebra is a central extension of $\vect(S^1)$.  As a vector space it is $\vir=\vect(S^1)\times i\mathbb{R}$ (the factor of $i$ and similar factors below are conventional and simplify some formulas).  So an element of $\vir$ is a pair, $(f(\theta)\partial_\theta,-ia)$, where $a\in \mathbb{R}$.  Elements of $\vir$ are added and subtracted in the obvious way.  What is nontrivial is the Lie bracket on $\vir$.  It has the form
\beq\label{eq:bracket}
[(f(\theta)\partial_\theta,-ia),(b(\theta)\partial_\theta,-ib)]
	=\left([f(\theta)\partial_\theta,g(\theta)\partial_\theta],\frac{-i}{48\pi}\omega(f(\theta)\partial_\theta,g(\theta)\partial_\theta)\right).
\eeq
The first entry on the rhs is the usual vector field commutator.  The function appearing in the second entry, $\omega(f(\theta)\partial_\theta,g(\theta)\partial_\theta)$, is a bilinear function of vector fields.  The form of $\omega$ is constrained by the requirement that the bracket \eqref{eq:bracket} is indeed a Lie bracket, and in particular that it satisfies the Jacobi identity.  This requirement is so severe that there is in fact a unique solution (up to rescalings),
\beq\label{eq:cocycle}
\omega(f(\theta)\partial_\theta,g(\theta)\partial_\theta)
	= 2\int_{S^1} f'(\theta)g''(\theta)\ud\theta,
\eeq
where $'\equiv \ud/\ud\theta$. Equations \eqref{eq:bracket} and \eqref{eq:cocycle} define the Lie bracket on the Virasoro algebra.

At this point it is helpful to introduce a basis for $\vir$ and express the Virasoro commutation relations in their usual form.  The usual basis elements ($m\in \mathbb{Z}$) are
\beq\label{eq:basis1}
v_m = \left(e^{im\theta}\partial_\theta,-\frac{i}{24}\delta_m^0\right).
\eeq
Any element of $\vir$ can be expressed as a linear combination of $v_m$.  The second entry on the rhs, $-i/24\delta_m^0$, is conventional and we will comment on its significance in a moment.  For now, observe that plugging the basis elements into the Lie bracket gives
\beq\label{eq:algebra}
i[v_m,v_n]=(m-n)v_{m+n} + \frac{1}{12}(m^3-m)\delta_{m+n}^0\mathcal{Z},
\eeq
where $\mathcal{Z}\equiv(0,-i)$.  These commutation relations are often presented as the definition of the Virasoro algebra.  The advantage of starting from the basis-independent presentation \eqref{eq:bracket}--\eqref{eq:cocycle} is that we are free to switch to bases other than the $v_m$.  This will become useful below.

The central extension appears in \eqref{eq:algebra} as a term proportional to $\mathcal{Z}$, with contributions linear and cubic in $m$.  The  linear in $m$ contribution could have been eliminated by setting the second entry of $v_m$ to zero in \eqref{eq:basis1}.  However, it is standard to include this term. It has the convenient effect of making the central term vanish for commutators involving only $v_0$ and $v_{\pm 1}$.  This will not be important for us, but we follow convention.  Note that the $m^3$ term cannot be eliminated by changing the basis elements.  
Non-trivial central extensions are classified by 2-cocycles on  $\vir$, which are elements of the second cohomology $H^2(\vir,\mathbb{R})$.  This is one-dimensional and generated by $\omega(f(\theta)\partial_\theta,g(\theta)\partial_\theta)$ \cite{khesin2008geometry}. The term linear in $m$ is a 2-coboundary and can be removed by performing a change of basis.

Returning to the general theory, consider the dual space, $\vir^*$, the space of linear functionals on $\vir$.  Its elements are pairs, $(u(\theta)d\theta^2,ic)$.  The pairing between $\vir$ and $\vir^*$ is
\beq\label{eq:pairing}
\langle (f(\theta)\partial_\theta,-ia),(u(\theta)d\theta^2,ic) \rangle = \int_{S^1} f(\theta) u(\theta) d\theta+ac.
\eeq
For each element of $\vir^*$, this pairing gives a linear map $\vir\rightarrow \mathbb{R}$.  This justifies our identification of $\vir^*$ with the dual space of $\vir$.  As before, note that $\vir$ and $\vir^*$ are isomorphic as vector spaces but they transform differently under diffeomorphisms.  Elements of $\vir$ are called adjoint vectors and elements of $\vir^*$ are called coadjoint vectors.  The adjoint action of $\vir$ on itself is just the Lie bracket \eqref{eq:bracket}:
\beq\label{eq:adjoint}
\ad_{(f(\theta)\partial_\theta,-ia)} (g(\theta)\partial_\theta,-ib)
	= \left([f(\theta)\partial_\theta,g(\theta)\partial_\theta],\frac{-i}{48\pi}\omega(f(\theta)\partial_\theta,g(\theta)\partial_\theta)\right).
\eeq
Note that $a$ and $b$ do not enter the rhs.  This reflects the fact that the extension is central, i.e., that $\mathbb{R}$ is in the center of $\vir$.  To streamline notation we will sometimes not write $a$ and $b$ explicitly on the lhs either.

The adjoint action can be transported to $\vir^*$ using the pairing \eqref{eq:pairing}.  This defines the coadjoint action,
\beq\label{eq:coadjoint}
\ad^*_{(f(\theta)\partial_\theta,-ia)}(u(\theta)d\theta^2,ic)
	= -\left(\left(u'f+2uf'-\frac{c}{24\pi}f'''\right)d\theta^2,0\right).
\eeq
The first thing to note is that the coadjoint action is different from the adjoint action \eqref{eq:adjoint}.  This is why the distinction between $\vir$ and $\vir^*$ is important: adjoint vectors and coadjoint vectors transform differently under infinitesimal diffeomorphisms (an infinitesimal diffeomorphism is a vector field).  In physical problems that arise  ``in the wild,'' it is not always obvious at first glance that a field is an element of $\vir^*$.  The way to check this is to see if the field transforms under diffeomorphisms according to the Virasoro coadjoint action \eqref{eq:coadjoint}.  In problems with hidden Virasoro symmetry, the third derivative in \eqref{eq:coadjoint} often manifests as a signal of the underlying symmetry.

The infinitesimal coadjoint action \eqref{eq:coadjoint} can be integrated to get an action of $\diff(S^1)$ on $\vir^*$.  The result is
\beq\label{eq:coadjoint2}
\Ad^*_{\eta^{-1}}(u(\theta)d\theta^2,ic) = \left(u(\eta)\cdot(\eta')^2 d\theta^2-\frac{c}{24\pi}\sch(\eta)d\theta^2,ic\right),
\eeq
where $\eta\in \diff(S^1)$ and $\sch(\eta)=(\eta' \eta'''-\tfrac{3}{2}(\eta'')^2)/(\eta')^2$ is its Schwarzian derivative.  Note that the central charge, $c$, is  invariant under the coadjoint action.  In other words, the central charge is constant on orbits of the coadjoint action.  When we speak of ``the'' central charge of a physical system, we mean the phase space of the system is a Virasoro coadjoint orbit with central charge $c$.

The dual of the Lie algebra, $\vir^*$, has a Poisson bracket called the Lie-Poisson bracket.  A Poisson bracket is an antisymmetric bilinear map satisfying Leibniz's rule and the Jacobi identity.  Let $F$ and $G$ be functions on $\vir^*$.  At a fixed point $\mu\equiv(u(\theta)d\theta^2,ic)\in \vir^*$, the Lie-Poisson bracket is
\beq\label{eq:pb}
\{F,G\}(\mu) = \bigg\langle \mu,\left[\frac{\delta F}{\delta \mu},\frac{\delta G}{\delta \mu}\right] \bigg\rangle.
\eeq
On the rhs, $\delta F/\delta \mu$ and $\delta G/\delta \mu$ are functional derivatives, considered as elements of $\vir$, and the bracket is the Lie bracket.  Equipped with the Lie-Poisson bracket, $\vir^*$ can serve as a phase space for Hamiltonian systems.  Let $H$ be a function on $\vir^*$, then Hamilton's equations are given by
\beq
\frac{\partial F}{\partial t} = \{F,H\}.
\eeq
Expand the lhs as $\partial F/\partial t = \langle \partial_t \mu, \delta F/\delta \mu \rangle$ and expand the rhs as
\beq\label{eq:LPbracket}
\{F,G\}(\mu)
	= \left\langle \mu,\left[\frac{\partial F}{\partial \mu},\frac{\partial H}{\partial \mu}\right]\right\rangle
	= -\left\langle \mu, \ad_{\delta H/\delta \mu} \frac{\delta F}{\delta \mu} \right\rangle
	= \left\langle \ad^*_{\delta H/\delta\mu }\mu, \frac{\delta F}{\delta \mu} \right\rangle.
\eeq
Comparing these expressions gives
\beq\label{eq:flow}
\frac{\partial \mu}{\partial t} = \ad^*_{\delta H/\delta \mu} \mu.
\eeq
Equations \eqref{eq:coadjoint} and \eqref{eq:flow} define a Hamiltonian flow on $\vir^*$ for each choice of $H$.  According to \eqref{eq:coadjoint}, the central charge is invariant under Hamiltonian flows; hence, the one-form $u(\theta)d\theta$ evolves according to 
\beq\label{eq:eom}
\frac{\partial u}{\partial t}= -u'f-2uf'+\frac{c}{24\pi}f''',
\eeq
where (abusing notation slightly) $f\equiv \delta H/\delta u$.  In applications, one often ``works backwards:'' given an equation of motion of the form \eqref{eq:eom}, one finds a Hamiltonian for which the equation can be realized as Hamiltonian flow on $\vir^*$.  For example, the Korteweg-de Vries (KdV) equation was discovered long ago as a model for one-dimensional fluid flow and only much later realized as a Hamiltonian flow on $\vir^*$ \cite{khesin2008geometry}.  In the next section, we will find equations governing the evolution of the boundary of three-dimensional asymptotically (A)dS spacetimes.  We will realize the boundary dynamics as Hamiltonian flow on $\vir^*$ by comparing the boundary's equations of motion with \eqref{eq:eom} and choosing $H$ appropriately.

\section{Anti-de Sitter}
\label{sec:ads}

The metric of three-dimensional anti-de Sitter (\ads) spacetime is
\beq\label{eq:gads}
ds^2 = \frac{\ell^2}{r^2}dr^2 - r^2 d\xp d\xm,
\eeq
where $x^\pm=t/\ell\pm\phi$ are null coordinates.
To study fluctuations about \eqref{eq:gads}, we introduce a family of metrics which approaches \eqref{eq:gads} asymptotically, as $r\rightarrow \infty$.  A seemingly reasonable ansatz for this family of metrics is
\beq\label{eq:ansatz0}
ds^2 = \frac{\ell^2}{r^2}dr^2 
	- e^{2\varphi}r^2 d\xp d\xm 
	+ e^{4\varphi} \gamma_{++} (d\xp)^2 
	+ e^{4\varphi} \gamma_{--} (d\xm)^2 
	+ (\dots),
\eeq
where $\varphi(x^+,x^-), \gamma_{++}(x^+,x^-),$ and $\gamma_{--}(x^+,x^-)$ are independent of $r$ and $(\dots)$ indicates subleading terms in $r^{-1}$.  This ansatz is not quite right because the curvature of \eqref{eq:ansatz0} falls off too slowly\footnote{Let $h_{\mu\nu}$ be the induced metric on the boundary (see below for formula).  The rhs of Einstein's equation gives $8\pi G\sqrt{-h}T^r_r=O(1)$ as $r\rightarrow \infty$, violating the requirement that the metric approaches vacuum AdS$_3$ asymptotically.} as $r\rightarrow \infty$.
To fix this, take instead
\beq\label{eq:ansatz1}
ds^2 = \frac{\ell^2}{r^2}dr^2 
		- (e^{2\varphi}r^2-2\gamma_{+-}) d\xp d\xm 
		+ e^{4\varphi} \gamma_{++} (d\xp)^2 
		+ e^{4\varphi} \gamma_{--} (d\xm)^2 
		+ (\dots),
\eeq
where $\gamma_{+-}=\ell^2 \partial_+ \partial_- \varphi$.  Equation \eqref{eq:ansatz1} is known to be the most general asymptotically-AdS metric in three dimensions \cite{Barnich:2010eb}.

The metric functions $\varphi, \gamma_{++}$, and $\gamma_{--}$ are not entirely arbitrary.  They satisfy two constraints coming from the $r\rightarrow\infty$ limit of the the $r\pm$-components of Einstein's equation.  Following the membrane paradigm, we regard these constraints as equations of motion for a dynamical system in 1+1 dimensions.  Comparing the constraints with \eqref{eq:eom} will allow us to realize this 1+1-dimensional dynamics as a Hamiltonian system on a coadjoint orbit of $\vir^* \times \vir^*$ with $c=3\ell/2G$.

\subsection{Boundary dynamics}

We fix a cutoff surface at large but finite $r=r_c$; this will serve as a proxy for the boundary of spacetime.    Ultimately we are interested in the $r_c\rightarrow \infty$ limit.  Let $n\equiv (r/\ell)\partial_r$ be the unit normal and $h_{\mu\nu}=g_{\mu\nu}-n_\mu n_\nu$ be the induced metric on the cutoff surface.  At the cutoff, the Einstein equations enforce the constraints\footnote{If one sets $\gamma_{\pm\pm}=0$, then the constraint may be interpreted as equations of motion for $\varphi$ in a flat $1+1$ spacetime. The dynamics of $\varphi$ is governed by Liouville theory \cite{Coussaert:1995zp,Cacciatori:2001un}. If instead we consider a non-zero $\gamma_{\pm\pm}$, then the boundary is curved and Liouville theory lives in a curved $1+1$ spacetime. Moreover, the $\pm\pm$ components of the Brown-York stress-tensor are proportional to the Liouville stress-tensor while the $+-$ components are proportional to the curvature of the boundary \cite{Balasubramanian:2002zh}.}
\begin{align}
\frac{1}{8\pi G}\sqrt{-h} G_{n+} 
	&= \frac{\partial}{\partial \xm} 
		\left[\frac{e^{4\varphi}\gamma_{++}}{8 \pi G\ell} 
		- \frac{\ell}{8\pi G}\partial_{+}^2 \varphi + \frac{\ell}{8\pi G}(\partial_+ \varphi)^2\right] = 0,\label{eq:constraint1}\\
\frac{1}{8\pi G}\sqrt{-h} G_{n-} 
	&= \frac{\partial}{\partial \xp} 
		\left[\frac{e^{4\varphi}\gamma_{--}}{8 \pi G\ell} 
		- \frac{\ell}{8\pi G}\partial_{-}^2 \varphi + \frac{\ell}{8\pi G}(\partial_- \varphi)^2 \right] = 0,\label{eq:constraint2}
\end{align}
where terms subleading in $1/r$ have been dropped.  In section \ref{sec:membrane}, we will explain the fluid interpretation of \eqref{eq:constraint1}-\eqref{eq:constraint2}, following the membrane paradigm.  The goal for the remainder of this section is to realize these equations as a Hamiltonian flow on $\vir^*\times \vir^*$.  

The last two terms on the rhs's of \eqref{eq:constraint1}-\eqref{eq:constraint2} are reminiscent of the Schwarzian derivatives in the transformation law \eqref{eq:coadjoint2} for Virasoro coadjoint vectors.  To make this correspondence precise, let $\partial_+ \eta_+ = e^{2\varphi}$ and  $\partial_- \eta_- = e^{2\varphi}$.  The Schwarzian derivative of $\eta_+$ is 
\beq
{\rm Sch}(\eta_+) 
	= \frac{(\partial_+ \eta_+)(\partial_+^3 \eta_+)-\tfrac{3}{2}(\partial_+^2 \eta_+)^2}{(\partial_+ \eta_+)^2}
	= - 2\partial_+^2 \varphi + 2(\partial_+ \varphi)^2.
\eeq
A similar equation holds for ${\rm Sch}(\eta_-)$.  Note that we always take ``spatial derivatives'' of $\eta_+$ with respect to  $x^+$ and ``spatial derivatives'' of $\eta_-$ with respect to $x^-$.  The constraint equations \eqref{eq:constraint1}-\eqref{eq:constraint2} can be written compactly as
\beq
\frac{\partial}{\partial x^\mp} \left(\Gamma_{\pm\pm}\cdot (\partial_\pm \eta_\pm)^2 - \frac{c}{24\pi}{\rm Sch}(\eta_\pm)\right) = 0,
\eeq
where $\Gamma_{\pm\pm}\equiv \gamma_{\pm\pm}/(8\pi G\ell)$ and $c=3\ell/2G$.  Comparing with \eqref{eq:coadjoint2}, we see that $\Gamma_{++}$ and $\Gamma_{--}$ transform as Virasoro coadjoint vectors with identical central charges, $c=3\ell/2G$.  The constraint equations become
\beq\label{eq:eom0}
\frac{\partial}{\partial x^\mp} \Ad^*_{\eta_{\pm}^{-1}} \left(\Gamma_{\pm\pm}(dx^{\pm})^2,ic\right) = 0.
\eeq
Using the ``product rule'' for the coadjoint action\footnote{See Proposition 9.3.8 of \cite{marsden2002introduction}.}, this becomes 
\beq\label{eq:eom1}
\frac{\partial}{\partial x^\mp} (\Gamma_{\pm\pm} (dx^\pm)^2,ic)  - \ad^*_{f_\pm}(\Gamma_{\pm\pm} (dx^\pm)^2,ic) = 0,
\eeq
where $f_\pm \equiv \partial_\mp \eta_\pm \in \lieg$ are the adjoint vectors corresponding to the diffeomorphisms $\eta_\pm$.  They can be thought of as the generators of an infinite dimensional generalization of translations along $x_\pm$.  From this, one can read off the equation of motion for $\Gamma_{++}$ which is 
\beq\label{eq:eompp}
\frac{\partial \Gamma_{++}}{\partial x^-}= -(\partial_+\Gamma_{++})f_+-2\Gamma_{++}(\partial_+ f_+) +\frac{c}{24\pi}\partial_+^3f_+.
\eeq
Notice that the dynamics of $\Gamma_{++}$ is precisely of the form \eqref{eq:eom}, with $x^-$ playing the role of ``time'' and $x^+$ playing the role of ``space.'' So the dynamics of $\Gamma_{++}$ can be realized as a Hamiltonian flow on a coadjoint orbit of $\vir^*$ with central charge $c=3\ell/2G$; all that remains is to specify a Hamiltonian, $H_+$, with $\delta H_+/\delta \Gamma_{++} = f_+ = \partial_- \eta_+$.  This can be achieved by defining
\beq
H_+ \equiv \int_{S^1} f_+ \Gamma_{++}  dx^+. 
\eeq
In the same way, we may regard the dynamics of $\Gamma_{--}$ as a Hamiltonian flow on a second copy of $\vir^*$, albeit with the roles of $x^+$ and $x^-$ interchanged. Since the dynamics of $\Gamma_{++}$ and $\Gamma_{--}$ are decoupled, one can slightly  abuse notation and consider a Hamiltonian, $H:\vir^*\times \vir^*\rightarrow \mathbb{R}$, for the combined system given by $H=H_+ + H_-$ where the first term evolves in ``time'' $x^-$  and the second one in ``time'' $x^+$.

At this point, we have constructed a Hamiltonian flow on $\vir^*\times \vir^*$ which may be studied as an interesting dynamical system in its own right, independent of its origins in ${\rm AdS}_3$ gravity.  The equation of motion \eqref{eq:eompp} is reminiscent of the Korteweg-de Vries equation, another example of a Hamiltonian flow on $\vir^*$ \cite{khesin2008geometry}.  The unusual feature of our Hamiltonian is the arbitrary function, $f_\pm(x^+,x^-)$.  In some sense, we really have a family of Hamiltonian systems parametrized by $f_\pm(x^+,x^-)$.

\subsection{Charge algebra}

Returning to the boundary equations of motion  \eqref{eq:eom0}, we have the conservation law
\beq\label{eq:charges}
\frac{\partial}{\partial x^-} \Xi_{++}(x^+,x^-) = 0,
\eeq
where $\Xi_{++}\equiv \Ad^*_{\eta_{+}^{-1}} \left(\Gamma_{++}(dx^{+})^2,ic\right)$ is an element of $\vir^*$ and we regard $x^-$ as time.  Now, $\vir^*$ is an infinite dimensional vector space, so  projecting \eqref{eq:charges} onto a basis for $\vir$ gives an infinite set of conserved charges.  The usual choice is
\beq\label{eq:adsbasis}
v^+_m=\left(e^{imx^+}\partial_+,-\frac{i}{24}\delta_m^0\right).
\eeq
Projecting onto this basis using \eqref{eq:pairing} gives an infinite set of conserved charges,
\beq
\mathcal{Q}^+_m \equiv \langle \Xi_{++}(x^+)  ,v_m^{+}(x^+) \rangle = {\rm const.}
\eeq
The Lie-Poisson bracket \eqref{eq:pb} of the charges is 
\begin{align}
i\{\mathcal{Q}_m^+,\mathcal{Q}_n^+\}(\Xi_{++})
	&= \langle \Xi_{++},i[v_m^+,v_n^+]\rangle ,\notag\\
	&= (m-n)\langle \Xi_{++},v_{m+n}\rangle + \frac{1}{12}m(m^2-1)\delta^0_{m+n}\langle(0,ic),(0,-i)\rangle,\notag\\
	&= (m-n)\mathcal{Q}^+_{m+n}+\frac{c}{12}m(m^2-1)\delta_{m+n}^0.
\end{align}
We get a second copy of this algebra by considering the boundary equation of motion $\partial_+ \Xi_{--}=0$ and regarding $x^+$ as ``time.'' In this case, the Lie-Poisson brackets of the corresponding charges, say $\mathcal{Q}^-_m$, are
\begin{align}
i\{\mathcal{Q}^-_m,\mathcal{Q}^-_n\}&=(m-n)\mathcal{Q}^-_{m+n}+\frac{c}{12}m(m^2-1)\delta_{m+n}^0, \\
i\{\mathcal{Q}^+_m,\mathcal{Q}^-_n\}&=0.
\end{align}
This is the asymptotic charge algebra of ${\rm AdS}_3$ gravity derived long ago by Brown and Henneaux \cite{Brown:1986nw}, with central charge $c=3\ell/2G$.  We have arrived at this result in a new way: by regarding the constraints on the boundary data as a 1+1-dimensional Hamiltonian system and using the corresponding Lie-Poisson bracket \eqref{eq:pb} on $\vir^* \times \vir^*$ to compute the charge algebra.  The simplicity of the Lie-Poisson bracket on $\vir^* \times \vir^*$ has helped to streamline the derivation.

\subsection{Membrane paradigm}
\label{sec:membrane}

The idea of regarding the constraints on the boundary data as a dynamical system in 1+1 dimensions is inspired by the black hole membrane paradigm.  In this section, we will use the relationship with the membrane paradigm to clarify the connection between gravity and fluid dynamics.  

Previously we introduced a cutoff surface at large but finite $r=r_c$, with induced metric $h_{\mu\nu}$ and unit normal $n$.  This is the analogue of the ``stretched horizon'' of the black hole membrane paradigm.  Following the membrane paradigm, we now assign the cutoff surface a stress-energy tensor, $t_{\mu\nu}$.  The stress-energy tensor is defined such that it terminates the gravitational field at the cutoff.  The Israel junction condition gives 
\beq
t_{\mu\nu} = -\frac{1}{8\pi G}\left(K_{\mu\nu}-\left(K-\frac{1}{\ell}\right)h_{\mu\nu}\right),
\eeq
where $K^\mu_\nu={h^\delta}_\nu \nabla_\delta n^\mu$ is the extrinsic curvature of the cutoff surface and $K$ is its trace. The term proportional to $1/\ell$ is a regularization term added in order to have a finite stress tensor as we approach the boundary\footnote{Note, our correction differs from \cite{Balasubramanian:1999re} by a minus sign because we use a different convention for defining the extrinsic curvature.} \cite{Balasubramanian:1999re}. Plugging in the metric \eqref{eq:ansatz1} gives
\beq \label{BYT}
t_{\pm\pm} =\frac{e^{4\varphi}\gamma_{\pm\pm}}{8\pi G \ell} = e^{4\varphi}\Gamma_{\pm\pm},
\eeq
for the diagonal components, while the off-diagonal ones are proportional to the Ricci scalar of the cutoff surface. Now, the variables $\Gamma_{\pm\pm}$ introduced earlier may be interpreted as energy densities.  They are the energy densities of the boundary stress-energy tensor as viewed by null observers with four-momenta $k_\pm=e^{-2\varphi}\partial_\pm$. It is interesting to look at the trace of the stress-energy tensor where we already see a hint of conformal invariance \cite{Balasubramanian:1999re}:
\beq
t^\text{reg}=\frac{c }{24 \pi} R_{2d}, \quad c=\frac{3\ell}{2G}  \ ,
\eeq
where $R_{2d}=8 e^{-2\varphi}\gamma_{+-}/(r \ell)^2$ is the Ricci scalar of the cutoff surface. This corresponds to the trace anomaly of a conformal field theory with central charge\footnote{Note that compared to \cite{Balasubramanian:1999re} we differ by a minus sign in the conformal anomaly. This is due to the fact that our conventions for the Riemann tensor differ by a minus sign.} $c=3\ell/2G$. This anomaly only exists for odd bulk dimensions. In these cases, a logarithmic divergence in the action appears and cannot be canceled by a polynomial counterterm without including a cutoff dependence \cite{Henningson:1998gx,Emparan:1999pm,deHaro:2000vlm,Skenderis:2000in,Balasubramanian:2001nb}. Given that the stress tensor measures the change in the action due to perturbations of the boundary, one finds this anomaly by looking at the trace of \eqref{BYT}.

Now, we move on to analyze the dynamics. Earlier we obtained equations of motion \eqref{eq:constraint1}-\eqref{eq:constraint2} for $\Gamma_{\pm\pm}=e^{-4\varphi}t_{\pm\pm}$ by projecting Einstein's equations onto the cutoff surface.  We obtain the same conservation laws by imposing conservation of $t_{\mu\nu}$:
\beq\label{eq:conservT}
\sqrt{-h}h_{a\mu}{t^{\mu\nu}}_{|\nu}=0,
\eeq
where the covariant derivative is ${t^{\mu\nu}}_{|\nu}=h^\delta_\nu\nabla_\delta t^{\mu\nu}$ and the index $a$ corresponds to $x_{\pm}$.  This means that the constraint equations \eqref{eq:constraint1}-\eqref{eq:constraint2} may be interpreted as energy conservation laws for the boundary stress-energy tensor.

Ordinary (inviscid, nonrelativistic) fluids are governed by the compressible Euler equations.  Mass conservation is described by the continuity equation
\beq\label{eq:cont}
\frac{\partial \rho}{\partial t} + \nabla \cdot (\rho v) = 0,
\eeq
where $\rho$ is the fluid's mass density and $v$ is its velocity.   Let $X$ and $x$ denote the positions of a fluid parcel at $t=0$ and time $t$, respectively.  These are called Lagrangian and Eulerian coordinates. Consider the diffeomorphism $\eta:X\rightarrow x$.  It is related to the fluid velocity by 
\beq
\frac{\partial \eta(X,t)}{\partial t} = v(\eta(X,t),t).
\eeq
In these variables, the continuity equation \eqref{eq:cont} is
\beq\label{eq:cont2}
\frac{\partial}{\partial t} \Ad^*_{\eta^{-1}} (\rho(x)dx) = 0,
\eeq
where $\Ad^*_{\eta^{-1}} (\rho(x)dx) = \eta^* (\rho(x)dx)$ is the adjoint action of $\diff(M)$ on $\Omega^n(M)$.  Here $M$ is the manifold on which the fluid is flowing and $\Omega^n(M)$ is the space of $n$-forms on $M$.

Recall that the dynamics governing the ${\rm AdS}_3$ boundary was described by
\beq\label{eq:eom0copy}
\frac{\partial}{\partial x^\mp} \Ad^*_{\eta_{\pm}^{-1}} \left(\Gamma_{\pm\pm}(dx^{\pm})^2,ic\right) = 0.
\eeq
Comparing \eqref{eq:cont2} and \eqref{eq:eom0copy}, we see that the maps $\eta_{\pm}$ are analogous to the map $\eta$ from Lagrangian to Eulerian coordinates.  In fact, the gravity equations \eqref{eq:eom0copy} are almost equivalent to two copies of the usual fluid continuity equation, with the only difference being that the dynamics takes place on $\vir^*$ rather than $\Omega^n(M)$.

\section{de Sitter}
The metric of three-dimensional de Sitter space (\ds) is
\beq\label{eq:gds}
ds^2 = -\frac{\ell^2}{r^2}dr^2 + r^2 d\tp d\tm,
\eeq
where $t^\pm=t/\ell \pm i\phi$.  This is related to the \ads\ metric \eqref{eq:gads} by the substitution $\ell\rightarrow -i\ell$, in which case $t^\pm \rightarrow i x^\pm$.  In this section, we extend our earlier discussion to asymptotically \ds\ spacetimes.  The calculations largely resemble the \ads\ case, so we will be concise.

To begin, we introduce a family of metrics which approaches \eqref{eq:gds} asymptotically, as $r\rightarrow \infty$:
\beq\label{eq:ansatz1dS}
ds^2 = -\frac{\ell^2}{r^2}dr^2 + (e^{2\varphi}r^2-2L_{+-}) d\tp d\tm - e^{4\varphi}L_{++} (d\tp)^2 - e^{4\varphi}L_{--} (d\tm)^2 + (\dots),
\eeq
where $L_{+-}\equiv\ell^2 \partial_+ \partial_- \varphi$. This is the \ds\ analogue of \eqref{eq:ansatz1}.  The metric functions  
$\varphi(t^+,t^-)$, $L_{++}(t^+,t^-)$, and $L_{--}(t^+,t^-)$ are independent of $r$.  Note that outside the cosmological horizon, $r$ is a timelike coordinate and $t$ is a spacelike coordinate. 

As before, we fix a cutoff surface at large but finite $r=r_c$.  This is a proxy for the ``boundary of spacetime'' at $r\rightarrow \infty$.  In the \ads\ case, the cutoff  was a timelike surface but now it is spacelike.  Let $n\equiv(r/\ell)\partial_r$ be the outward pointing unit normal and $h_{\mu\nu}=g_{\mu\nu}+n_\mu n_\nu$ be the induced metric on the cutoff surface.  The Einstein equations enforce the constraints
\begin{align}
\frac{1}{8\pi G}\sqrt{h} G_{n+} 
	& = \frac{\partial}{\partial \tm} 
	\left[\frac{e^{4\varphi} L_{++}}{8 \pi G\ell}
	- \frac{\ell}{8\pi G}\partial_{+}^2 \varphi + \frac{\ell}{8\pi G}(\partial_+ \varphi)^2\right] = 0,\label{eq:constraint1dS}\\
\frac{1}{8\pi G}\sqrt{h} G_{n-} 
	&= \frac{\partial}{\partial \tp} 
	\left[\frac{e^{4\varphi} L_{--}}{8 \pi G\ell} 
	- \frac{\ell}{8\pi G}\partial_{-}^2 \varphi + \frac{\ell}{8\pi G}(\partial_- \varphi)^2 \right] = 0.\label{eq:constraint2dS}
\end{align}
Following the \ads\ example, we set $\mathcal{L}_{\pm\pm}\equiv L_{\pm\pm}/(8\pi G \ell)$, $\partial_+ \eta_+ = e^{2\varphi}$, and $\partial_- \eta_- = e^{2\varphi}$ (in this section, $\partial_{\pm} = \partial_{t^{\pm}}$).  Further define
\beq \label{codS}
\Theta_{\pm\pm}\equiv\Ad^*_{\eta_\pm^{-1}}(\mathcal{L}_{\pm\pm}(\ud t^\pm)^2, ic).
\eeq
The constraint equations \eqref{eq:constraint1dS}-\eqref{eq:constraint2dS} become simply
\beq
\partial_\mp \Theta_{\pm\pm} = 0.
\eeq
So we are in the same situation as before.  As in the \ads\ case, the constraints give a Hamiltonian system on $\vir^*\times \vir^*$ with identical central charges $c=3\ell/2G$.

\subsection{Conserved quantities}

The conserved quantities, $\Theta_{\pm\pm}$, are elements of $\vir^*\times \vir^*$.  As before, we may project the conserved quantities onto a basis of $\vir\times\vir$.  In the present case, a natural basis for the left Virasoro is 
\begin{align}
l_m^+ &= \left(e^{-mt_+}\partial_{+},\frac{1}{24}\delta^0_m\right),
\end{align}
and the Lie brackets are
\beq
[l_m^+,l_n^+] = (m-n) l_{m+n}^+ + \frac{1}{12}(m^3-m)\delta_{m+n}^0\mathcal{Z}',
\eeq
where $\mathcal{Z}'=(0,1)$.  The conserved quantities are 
\beq
L_m^+ \equiv \langle l_m^+(t^+),-i\Theta_{++}(t^+) \rangle = {\rm const.},
\eeq
and the Lie-Poisson bracket is
\beq
\{L_m^+,L_n^+\} (-i\Theta_{++})= (m-n)L_{m+n}^+ + \frac{c}{12}(m^3-m)\delta_{m+n}^0.
\eeq
The renormalized Brown-York stress energy tensor is now
\beq
t^{\rm dS}_{\mu\nu} = -\frac{1}{8\pi G}\left(K_{\mu\nu}-\left(K-\frac{1}{\ell}\right)h_{\mu\nu}\right),
\eeq
whose diagonal components are 
\beq
t^{\rm dS}_{\pm\pm} =-e^{4\varphi} \mathcal{L}_{\pm\pm}. \label{tds}
\eeq

\subsection{Example: Kerr-${\rm dS}_3$}

To gain some intuition into the meaning of the $L_{\pm\pm}$ we can look at an example of an asymptotically dS$_3$ spacetime. Although there are no black hole solutions in dS$_3$, there are conical defects like the Kerr-dS$_3$ spacetime \cite{Deser:1983nh,Park:1998qk,Banados:1998tb}. The metric for Kerr-dS$_3$ is
\begin{align}
ds^2	=	&-\left(8GM-\frac{r^2}{\ell^2}+\frac{(8GJ)^2}{4r^2}\right)dt^2
	+\left(8GM-\frac{r^2}{\ell^2}+\frac{(8GJ)^2}{4r^2}\right)^{-1}dr^2\notag\\
	&+r^2\left(-\frac{8GJ}{2r^2}dt+d\phi\right)^2.
\end{align}
It is a quotient of dS$_3$ by a discrete group \cite{Balasubramanian:2001nb} similar to the BTZ black hole \cite{Banados:1992wn}. Rewriting this metric in $t_\pm$ coordinates and taking $t_\pm\rightarrow (1+2GM \ell^2/r^2)t_\pm$ we find that at large $r$,
\beq
ds^2 = -\frac{\ell^2}{r^2}dr^2 +r^2 d\tp d\tm- 2G\ell(\ell M+i J ) (d\tp)^2 - 2G\ell(\ell M-i J) (d\tm)^2 + (\dots).
\eeq
Comparing with \eqref{eq:ansatz1dS}, we find that the mass and angular momentum are 
\beq
M=\frac{L_{++}+L_{--}}{4G\ell^2} \ , \quad J=i\, \frac{L_{++}-L_{--}}{4G\ell}.
\eeq
Note that the $L_{\pm\pm}$ are complex valued.

\section{Discussion}

We analyzed the dynamics of the asymptotic data of \ds\ and \ads\ by realizing Einstein's equations as a 1+1-dimensional Hamiltonian system. We constructed an infinite set of conserved charges by projecting the conservation laws onto a basis for $\vir$ and we computed the asymptotic charge algebra using the Lie-Poisson bracket.  The Hamiltonian flow on $\mathfrak{g}^*$ is restricted to coadjoint orbits (c.f. eq. \ref{eq:eom1}), which are symplectic manifolds, so they are natural phase spaces for classical dynamics. Moreover, coadjoint orbits of $\mathfrak{g}^*$ can be identified with representations of the group $G$ according to Kirillov's orbit method \cite{kirillov2004lectures}. So this formulation of the classical theory can give insights into its quantization, see \cite{Witten:1987ty,Alekseev:1988ce,Alekseev:1988vx,Verlinde:1989hv,Alekseev:1990mp,Alekseev:2018ful} for some approaches.

As we mentioned at the outset, the subject of asymptotic symmetries has a wide range of applications, and our results give a new perspective on these applications.  For example, in cosmology, the change in the curvature perturbation, $\zeta$, produced by an asymptotic symmetry transformation of \ds\ is identified with adiabatic modes of $\zeta$ \cite{Hinterbichler:2016pzn}.  The relation to the membrane paradigm described in the present paper allows one to interpret the perturbations produced by adiabatic modes as changes in the energy density of the boundary fluid as observed by a null observer (c.f. eq. \ref{tds}).  Mathematically, they are paths along coadjoint orbits of $\vir^*\times \vir^*$.

\acknowledgments

It is a pleasure to thank Justin Khoury for helpful discussions. MCG is supported in part by US Department of Energy (HEP) Award de-sc0013528.  RFP is supported by a Prize Postdoctoral Fellowship in the Natural Sciences at Columbia University and by Simons Foundation Award Number 555117.  
\bibliography{msv5}
 
\end{document}